\documentclass[showpacs,preprintnumbers,amsmath,amssymb]{revtex4}
\usepackage{epsf}
\usepackage{amssymb}
\usepackage{amsfonts}
\usepackage{latexsym}
\usepackage{mathrsfs}
\usepackage{graphicx,epsf}
\def\be{\begin{equation}}
\def\ee{\end{equation}}
\def\bea{\begin{eqnarray}}
\def\eea{\end{eqnarray}}

\def\vphi{\varphi}

\def\nn{\nonumber}

\newcommand{\ik}[0]{\not {\! d}^{\,3}k}
\newcommand{\delbar}[1]{\not\!\delta^{\,3}(\bf{#1})}

\begin{document}
%
\title{Non-Gaussian perturbations from multi-field inflation}
\author{Laura E.~Allen, Sujata Gupta and David Wands}
%
%
%
\affiliation{
Institute of Cosmology and Gravitation, University of Portsmouth,
Portsmouth~PO1~2EG, United Kingdom}
%
%
%
\begin{abstract}

We show how primordial bispectrum of density perturbations from inflation may be characterised in terms of manifestly gauge-invariant cosmological perturbations at second order. The primordial metric perturbation, $\zeta$, describing the perturbed expansion of uniform-density hypersurfaces on large scales is related to scalar field perturbations on unperturbed (spatially-flat) hypersurfaces at first- and second-order. The bispectrum of the metric perturbation is thus composed of (i) a local contribution due to the second-order gauge-transformation, and (ii) the instrinsic bispectrum of the field perturbations on spatially flat hypersurfaces. We generalise previous results to allow for scale-dependence of the scalar field power spectra and correlations that can develop between fields on super-Hubble scales.

\end{abstract}

\pacs{98.80.Cq \hfill astro-ph/0509719 v2}

\maketitle


\section{Introduction}

Inhomogeneities in the distribution of matter and radiation in the
universe today provide valuable information about the dynamical
history and physical processes in the very early Universe. In
particular temperature anisotropies in the cosmic microwave background
(CMB) sky provide detailed information about the primordial density
perturbations at the time of last scattering of the CMB photons.
At present the simplest explanation for the origin of these
perturbations seems to be that they originated as quantum fluctuations
in one or more light fields during a period of inflation in the very
early Universe \cite{LiddleLyth,RMP}. These vacuum fluctuations on small (sub-Hubble) scales
were stretched to large (super-Hubble) scales by the accelerated
expansion, where they subsequently evolved as effectively classical
perturbations.

Vacuum fluctuations of massless scalar fields in a de Sitter geometry
enter a squeezed state on scales larger than the Hubble length where
the decaying mode can be neglected. Thus they can be treated as an
effectively classical distribution on large scales with Gaussian
statistics. For a Gaussian random field the power spectrum is sufficient to describe
all the statistical properties of the distribution. But
non-linearities in the evolution of the initial fluctuations will lead
to a non-Gaussian distribution.
Until recently studies of non-Gaussian perturbations from inflation were largely restricted to self-interactions of scalar fields \cite{Hodges,Falk,BernardeauUzan} neglecting non-linear gravity. Gravitational effects can be included in a stochastic approach using local equations to study the evolution of the coarse-grained inflaton field on long wavelengths \cite{Gangui,Gupta,Rigopoulos2,Hattori}. 

Non-Gaussianities inevitably arise at second-order in cosmological perturbation theory but the nature of the non-Gaussianity will depend upon the choice of variables \cite{LRa}. Acquaviva et al \cite{Acquaviva} constructed a gauge-invariant quantity at second-order on large scales, but its physical interpretation was left open and it was only approximately constant during slow-roll inflation \cite{Vernizzi,LRa}. Maldacena \cite{Maldacena} gave an analysis of second-order perturbations from slow-roll inflation working in terms of the field perturbation on unperturbed spatial hypersurfaces and relating these to the curvature perturbations on uniform field hypersurfaces during single field inflation. Both these quantities have a clear physical definition and hence are implicitly gauge-invariant. He showed that there is a curvature perturbation, which becomes constant on large scales and whose bispectrum for squeezed triangles ($k_1\ll k_2,k_3$) is proportional to the tilt of the scalar power spectrum, and hence small.
Maldacena's result has subsequently been verified by a number of authors \cite{Gruzinov,Creminelli,Seery1,Rigopoulos3,Seery2}, going beyond slow-roll \cite{Creminelli}, and extending it to loop corrections \cite{Weinberg} and non-minimally coupled fields \cite{Seery1}.
For a recent review see Ref.~\cite{review}.

Only multiple field models of inflation seem likely to generate significant non-Gaussianity in the density perturbation \cite{LindeMukhanov,BernardeauUzan,LUW,Zal,Seery2,Rigopoulos4,LRb,LZ}. In particular Lyth and Rodriguez \cite{LRb} have recently emphasized that the non-Gaussianity of the curvature perturbation may be simply written down in the so-called ``$\delta N$-formalism'' \cite{Starobinsky,SasakiStewart,LW03,LMS}
where one uses the ``separate universes'' approach \cite{WMLL} to follow the non-linear evolution on super-Hubble scales in terms of locally homogeneous solutions.

In this paper we show how the non-linear evolution of primordial
perturbations from inflation may be characterised in terms of
manifestly gauge-invariant perturbations at second-order. We adopt the formalism of Malik and Wands \cite{MW} who showed that there is a conserved curvature perturbation at second-order for adiabatic density perturbations on long wavelengths. 
In section~II we define the gauge-invariant curvature perturbation at second-order and give the leading order expressions for the power spectrum and bispectrum. In section~III we specialise to the case of adiabatic perturbations on large scales from single-field inflation. We relate the conserved curvature perturbation on large scales to the gauge-invariant field perturbations on spatially flat hypersurfaces during slow-roll inflation, consistent with the $\delta N$-formalism. In particular we distinguish the purely local form of the bispectrum generated by the second-order gauge transformation from the intrinsic bispectrum of the field perturbations on unperturbed spatial hypersurfaces. We use a simple argument to estimate the bispectrum of field due to large-scale variations in the local Hubble rate during inflation, and hence recover Maldacena's result for the non-linearity of the curvature perturbation. We generalise these results to the multiple-field case in Section~IV, allowing for scale dependence of the power spectra and cross-correlations between the fields which may develop on large scales. These results can be simplified by decomposing the field perturbations into an inflaton component (along the instantaneous background trajectory) and orthogonal isocurvature perturbations \cite{Gordon}. We conclude in Section~V.

\section{Gauge-invariant Perturbations}

Discussions of relativistic perturbations have historically been
plagued by gauge-dependence of apparently physical quantities such
as the density perturbation. For instance, in a spatially
homogeneous background where the density evolves in time, the
inhomogeneous density perturbation depends on the choice of time
coordinate at first-order, and spatial coordinates at second-order.
This ``gauge-problem'' can be avoided by using a physical choice
of gauge and building gauge-invariant definitions of the
perturbations in that gauge. Thus Bardeen constructed
gauge-invariant cosmological perturbations at first-order
\cite{Bardeen80} and Malik and Wands have shown how to construct
gauge-invariant cosmological perturbations at second-order \cite{MW}.
This still leaves a number of different possible gauge-invariant
definitions of the density perturbation, for example, corresponding to
different physical gauge choices, though the perturbations are
nonetheless gauge-invariant.


We will split any scalar $\phi$ into a spatially homogeneous
background and first- and second-order inhomogeneities:
\begin{equation}
\phi(t,x) = \phi_0(t) + \delta_1\phi(t,x) +\frac12 \delta_2\phi(t,x) \,.
\end{equation}
The split between first and second order inhomogeneities for any given
initial perturbation is arbitrary, so we can use this freedom to
specify that the first-order perturbations are Gaussian random fields
and any non-Gaussianity is described by the second-order perturbation.
In particular we will take the Gaussian part of the perturbations to
originate from free field fluctuations at early times during
inflation, and assume the second-order perturbations are vanishing in
this early time limit.

Under a temporal gauge transformation,
$\alpha=\alpha_1+(\alpha_2/2)$, a scalar $\phi$ transforms at first-
and second-order as
\begin{eqnarray}
 \delta_1\phi &\to& \delta_1\phi + \dot\phi_0\alpha_1 \,,\\
 \delta_2\phi &\to& \delta_2\phi + \dot\phi_0\alpha_2 + \partial_t \left(
 \dot\phi_0\alpha_1+2\delta_1\phi \right) \alpha_1
  \,.
 \end{eqnarray}
We will adopt the method of Malik and Wands \cite{MW} to define
gauge-invariant cosmological perturbations up to second order by
constructing a physical perturbation in a physically defined
coordinate system.

There are (at least) two different ways of writing the perturbed line
element on any spatial hypersurfaces:
\begin{eqnarray}
ds^2 &=& a^2(t) \left[ (1-2\psi)\delta_{ij}dx^i dx^j + 2 E_{,ij} +
  2F_{(i,j)} + h_{ij} \right]
 \,, \\
&=& a^2(t) \left[ e^{2\delta N}\delta_{ij}dx^i dx^j + 2 E_{,ij}  +
  2F_{(i,j)} + h_{ij} \right]
 \,,
\end{eqnarray}
where $F_i$ is the divergence-free, vector perturbation, and $h_{ij}$
is the transverse and trace-free tensor perturbation. One may refer to
$\psi$ as the curvature perturbation and $\delta N$ as the perturbed
expansion, though the two are trivially related as $\psi=(e^{2\delta
  N}-1)/2$. To first and second order we write
\begin{eqnarray}
\psi &=& \psi_1 + \frac12\psi_2  \,,\\
\delta N &=& \delta_1 N + \frac12 \delta_2 N \,,
\end{eqnarray}
and hence
\begin{eqnarray}
\psi_1 &=& - \delta_1 N \,,\\
\label{diff2}
 \psi_2 &=& - \delta_2 N - 2(\delta_1 N)^2 \,.
\end{eqnarray}

The curvature perturbation, $\psi$ or $\delta N$, on uniform-density
hypersurfaces is non-linearly conserved on large scales (where
gradient terms can be neglected) for adiabatic perturbations as a
direct consequence of local energy conservation
\cite{WMLL,LW03,MW,LMS}, (see also \cite{Rigopoulos1,LangloisVernizzi}). 
Thus we will be particularly interested in
calculating this perturbation.

Most authors \cite{Maldacena,LW03,Zaldarriaga,LMS} have worked in terms
of $\delta N$ on uniform-density hypersurfaces, which we denote by
\begin{eqnarray}
\label{zeta1}
 \zeta_1 &\equiv& \delta_1N - \frac{H}{\dot\rho_0}\delta_1\rho
 \,, \\
 \label{zeta2}
 \zeta_2 &\equiv& \delta_2 N  -
\frac{H}{\dot\rho_0}\delta_2\rho + \left(
\frac{H\delta_1\rho}{\dot\rho_0} \right) \left[ 2\left(
\frac{\dot{\delta_1\rho}}{\dot\rho_0} - \frac{\dot{\delta_1N}}{H}
\right) + \left( \frac{\dot{H}}{H} -\frac{\ddot\rho_0}{\dot\rho_0}
\right) \frac{\delta_1\rho}{\dot\rho_0} \right]
\,.
\end{eqnarray}

This definition leaves a residual dependence at second order on the
choice of spatial gauge, i.e., the choice of spatial coordinates
threading the uniform-density hypersurfaces. This can be eliminated
by a physical choice of spatial gauge at first-order \cite{MW}, such
as working with worldlines orthogonal to the uniform-density
hypersurfaces. However in what follows we will consider
perturbations on large scales where in any case the residual spatial
gauge dependence becomes negligible \cite{LW03}.

Malik and Wands \cite{MW} gave a gauge-invariant definition of the
curvature perturbation at second order on uniform-density
hypersurfaces, $\tilde\psi|_\rho$, and showed this is conserved for
adiabatic perturbations on large scales. This is identical to
$-\zeta$ defined above at first-order,
$\tilde\psi_1|_\rho=-\zeta_1$, but differs at second order simply
due to the difference in choice of curvature or expansion
perturbation (\ref{diff2}). Thus we have \cite{LRa}
\begin{equation}
\tilde\psi_2|_\rho = - (\zeta_2 + 2\zeta_1^2) \,.
\end{equation}
Hence both $\zeta$ and $\tilde\psi|_\rho$ are conserved at second
order for adiabatic perturbations on large scales.

For a stochastic random field, as would be expected to arise from
vacuum fluctuatuations during inflation in the early universe, the
perturbations are most usefully described in terms of their power
spectrum, and higher-order moments, in Fourier space. At linear order
the Fourier coefficients of $\zeta$ are defined by the same equation
(\ref{zeta1}) as in real space, while at second order we have
\begin{equation}
 \label{zeta2k}
 \zeta_{2,{\bf k}} \equiv \delta_2 N_{\bf k} -
  \frac{H}{\dot\rho_0}\delta_2\rho_{\bf k} + \left(\left(
\frac{H\delta_1\rho}{\dot\rho_0} \right) * \left[ 2\left(
\frac{\dot{\delta_1\rho}}{\dot\rho_0} - \frac{\dot{\delta_1 N}}{H}
\right) + \left( \frac{\dot{H}}{H} -\frac{\ddot\rho_0}{\dot\rho_0}
\right) \frac{\delta_1\rho}{\dot\rho_0} \right] \right)_{\bf k} \,,
\end{equation}
where ``$*$'' represents a convolution
\begin{equation}
\left( f*g \right) ({\bf k})
 \equiv \int^\infty_{-\infty}f({\bf k}-{\bf k'})\,g({\bf k'})\,\ik' \,,
\end{equation}
where $\ik$ represents $d^{\,3}k/(2\pi)^3$. We will be using the corresponding notation $\delbar{\bf{k}}$ to represent $(2\pi)^3\delta^{\,3}(\bf{k})$.

The power spectrum of the perturbation is given by
\begin{equation}
 P_\zeta({\bf k}) \delbar{{\bf k}+{\bf k_1}}
  \equiv
 \langle\zeta_{{\bf k}}\,\zeta_{{\bf k_1}}\rangle
  \,,
\label{defpowerspectrum}
\end{equation}
For an isotropic distribution the power spectrum is a function solely
of $k\equiv |{\bf k}|$.
Note that the variance per logarithmic interval in $k$-space is given by
\begin{equation}
{\cal P}_\zeta (k) = \frac{4\pi k^3}{(2\pi)^3} P_\zeta(k) \,.
\end{equation}
The bispectrum of the curvature perturbation is
\begin{equation}
 B_\zeta({\bf k_1},{\bf k_2})
 \delbar{{\bf k_1}+{\bf k_2}+{\bf k_3}}
 \equiv
\langle\zeta_{{\bf k_1}}\,\zeta_{{\bf k_2}}\,
 \zeta_{{\bf k_3}}\rangle \,,
 \label{defbispectrum}
\end{equation}
which by the requirement of statistical isotropy is non-zero only for
a triangle of modes, ${\bf k_1}+{\bf k_2}+{\bf k_3}=0$.

The bispectrum is zero for Gaussian perturbations and hence to leading
order we have
\begin{eqnarray}
 \label{defP}
 P_\zeta(k) \delbar{{\bf k}+{\bf k_1}}
  &=&
 \langle\zeta_{1,{\bf k}}\,\zeta_{1,{\bf k_1}}\rangle
  \,, \\
\label{defB}
 B_\zeta({\bf k_1},{\bf k_2})
 \delbar{{\bf k_1}+{\bf k_2}+{\bf k_3}}
 &=&
\frac12 \left[
 \langle\zeta_{1,{\bf k}}\,\zeta_{1,{\bf k'}}\, \zeta_{2,{\bf k''}}\rangle
\right. \nn\\
&& \left. +  \langle\zeta_{1,{\bf k}}\,\zeta_{2,{\bf k'}}\, \zeta_{1,{\bf k''}}\rangle
 +  \langle\zeta_{2,{\bf k}}\,\zeta_{1,{\bf k'}}\, \zeta_{1,{\bf k''}}\rangle
\right]
 \,.
\end{eqnarray}

If the second order part of the curvature perturbation can be
written purely in terms of the square of the first order part in
real space
\begin{equation}
 \label{deffnl}
 \zeta_{2}= - (6f_{NL}/5)\zeta_{1}^2\,,
\end{equation}
where $f_{NL}$ is a function of background parameters
then the bispectrum will be given by
\be
 B_\zeta(k_1,k_2,k_3)= - \frac{6}{5}f_{NL}
(P_\zeta(k_1)P_\zeta(k_2)+P_\zeta(k_1)P_\zeta(k_3)+P_\zeta(k_2)P_\zeta(k_3))\,,
\label{localB}
 \ee
calculated for closed triangles, ${\bf k_1}+{\bf k_2}+{\bf k_3}=0$. $f_{NL}$ is then a quantifier of this form of ``local'' non-Gaussianity and is called the non-linearity parameter.

\section{Single-field inflation}

We will now calculate the bispectrum for the scalar field
perturbations on large-scales in single-field models of inflation,
and hence reproduce the bispectrum for the large-scale curvature
perturbation, $\zeta$, \cite{Maldacena}.

\subsection{Curvature perturbations from inflaton perturbations}

During single field inflation, driven by field $\vphi$, density perturbations become adiabatic
in the large-scale limit at first- \cite{Gordon} or second-order
\cite{Vernizzi}. An important consequence is that $\zeta$ is
conserved and the uniform-density hypersurfaces coincide with
uniform-field hypersurfaces on large scales. Thus we have
\begin{eqnarray}
 \label{defzeta1phi}
 \zeta_1 &=& \delta_1 N - \frac{H}{\dot\vphi}\delta_1\vphi
 \,, \\
\label{defzeta2phi}
 \zeta_2 &=& \delta_2 N -
\frac{H}{\dot\vphi}\delta_2\vphi + \left(
\frac{H\delta_1\vphi}{\dot\vphi} \right) \left[
 2\left( \frac{\dot{\delta_1\vphi}}{\dot\vphi} -
 \frac{\dot{\delta_1 N}}{H} \right)
  + \left(
\frac{\dot{H}}{H} -\frac{\ddot\vphi}{\dot\vphi} \right)
\frac{\delta_1\vphi}{\dot\vphi} \right]
\,.
\end{eqnarray}

These expansion perturbations on uniform-field hypersurfaces are
closely related to the field perturbations on uniform-expansion
hypersurfaces, which have the gauge-invariant definitions
 \bea
 \label{defQ1}
\delta_1Q &\equiv& \delta_1\vphi - \frac{\dot\vphi}{H}\delta_1 N\,,\\
 \label{defQ2}
\delta_2Q &\equiv&  \delta_2\vphi - \frac{\dot\vphi}{H}\delta_2 N +
\left( \frac{\dot\vphi\delta_1 N}{H} \right) \left[ 2\left(
\frac{\dot{\delta_1 N}}{H} - \frac{\dot{\delta_1\vphi}}{\dot\vphi}
\right)
 + \left( \frac{\ddot\vphi}{\dot\vphi} - \frac{\dot{H}}{H} \right)
 \left( \frac{\delta_1 N}{H} \right) \right]
 \,.
 \eea
At first-order $\delta_1 Q$ is the usual Mukhanov-Sasaki variable \cite{Mukhanov,Sasaki}
used to quantise the semi-classical linear perturbations during
inflation, and $\delta_2 Q$ is its natural extension to second-order
\cite{Maldacena,MW}.

We can re-write the metric perturbation $\zeta$ in
Eqs.~(\ref{defzeta1phi}) and~(\ref{defzeta2phi}) somewhat more
succinctly in terms of these gauge-invariant field perturbations, to
give (see also Ref.\cite{Malik05})
\begin{eqnarray}
 \label{defzeta1Q}
 \zeta_1 &=& - \frac{H}{\dot\vphi}\delta_1Q \,, \\
\label{defzeta2Q}
 \zeta_2 &=&  -\frac{H}{\dot\varphi}\delta_2Q
  - \left(\frac{\dot H}{H^2}-\frac{\ddot\vphi}{H\dot\vphi}\right)
  \left(\frac{H\delta_1Q}{\dot\vphi}\right)^2
  -2\dot\zeta_1 \frac{\delta_1\vphi}{\dot\vphi} \,,
\end{eqnarray}
where we have used
\begin{equation}
 \frac{\dot{\delta_1 N}}{H} -
  \frac{\dot{\delta_1\vphi}}{\dot\vphi} =
 \frac{\dot\zeta_1}{H} + \left( \frac{\dot{H}}{H^2} -
 \frac{\ddot\vphi}{H\dot\vphi} \right)
 \frac{\delta_1\vphi}{\dot\vphi}
 \,.
 \end{equation}
Introducing the slow-roll parameters
 \bea
 \label{defepsilon}
\epsilon&\equiv&-\frac{\dot H}{H^2}\,,\\
\eta&\equiv&-\left(\frac{\dot H}{H^2}+\frac{\ddot
\vphi}{H\dot\vphi}\right)\,,
 \eea
and using the fact that $\dot\zeta_1=0$ for adiabatic perturbations
on large scales, we finally have
\begin{equation}
\label{defzeta2Q2}
 \zeta_2 =  -\frac{H}{\dot\varphi}\delta_2Q
  + \left(2\epsilon-\eta\right)
  \left(\frac{H\delta_1Q}{\dot\vphi}\right)^2
  \,.
\end{equation}

We can show that the non-linear adiabatic curvature perturbation on large
scales during inflation can simply be written as the perturbed
expansion history, $N=\int Hdt$, due to the gauge-invariant field
perturbations \cite{LW03,LRb}
\begin{equation}
 \label{defzeta}
\zeta = \frac{dN}{d\vphi} \delta Q + \frac12 \frac{d^2N}{d\vphi^2}
(\delta Q)^2 + \ldots \,.
\end{equation}
This is a non-linear extension of the $\delta N$-formalism \cite{Starobinsky,SasakiStewart} for calculating perturbations from inflation.
Hence to first and second order we have
\begin{eqnarray}
\label{defzeta1N}
 \zeta_1 &=& \frac{dN}{d\vphi} \delta_1 Q \,,\\
\label{defzeta2N}
 \zeta_2 &=& \frac{dN}{d\vphi} \delta_2 Q +
\frac{d^2N}{d\vphi^2} \delta_1 Q^2\,,
\end{eqnarray}
where
\begin{eqnarray}
 \label{d1N}
\frac{dN}{d\vphi} &=& - \frac{H}{\dot\vphi} \,,\\
 \label{d2N}
\frac{d^2N}{d\vphi^2} &=&
 - \left(\frac{\dot H}{H^2}-\frac{\ddot\vphi}{H\dot\vphi}\right)
  \left( \frac{H}{\dot\vphi} \right)^2
 = (2\epsilon-\eta) \left( \frac{H}{\dot\vphi} \right)^2 \,.
 \end{eqnarray}

\subsection{Non-Gaussianity of field perturbations}

During slow-roll inflation the initial state of the scalar field
perturbations can be set by imposing the Minkowski vacuum state on
small scales (much smaller than the cosmological Hubble scale,
$k/a\gg H$) for the scalar field perturbations on spatially-flat
hypersurfaces, $\delta N=0$. For any weakly-coupled scalar field
this gives an effectively Gaussian distribution of initial field
fluctuations. For any light field (with effective mass less than the
Hubble scale) living in a de Sitter spacetime, the Hubble damping
leads to a squeezed state on scales larger than the Hubble length
and hence we can treat the field as a classical random field with
\begin{equation}
P_Q( k_*) =  \frac{H^2}{2 k_*^3} ,
\label{Powerphi}
\end{equation}
and hence ${\cal P}_\zeta(k_*)=(H/2\pi)^2$, where a $*$ subscript
denotes Hubble-crossing, $k_*=aH$.

We will give a simple extension of the argument due to Maldacena
\cite{Maldacena,Creminelli} to calculate the amplitude
of the second-order (non-Gaussian) contribution to the scalar field
perturbation due to gravitational coupling at Hubble-exit.

It will be convenient to write the first-order field perturbation at
Hubble exit as 
\begin{equation}
\delta_1 Q ({\bf k_*}) = \frac{H}{2\pi} \hat{e}_{\bf k_*} \,.
\end{equation}
where $\hat{e}_{\bf k}$ is a classical Gaussian random variable with
unit variance:
\begin{equation}
\langle \hat{e}_{\bf k} \hat{e}_{\bf k'} \rangle = \frac{\delbar{{\bf k}+{\bf
  k'}}}{4\pi k^3}
 \,.
\label{thetaeqn}
\end{equation}

However at second order there is a local correction to the amplitude
of vacuum fluctuations at Hubble exit 
due to first-order perturbations in the local Hubble rate,
$\tilde{H}$. 
This is determined by the local scalar field
value due to longer wavelength modes that have already left the
horizon
\begin{equation}
\tilde{H} = H(\vphi_0) + H'(\vphi_0) \int_0^{k_c} \ik \,
\delta_1Q_{\bf k} \,.
\end{equation}
where $k_c < k_*$ is a cut-off wave-number.

Thus for $k_*\gg k_c$ we have upto second order
\begin{equation}
\delta Q ({\bf k_*}) = \frac{{H}}{2\pi} \hat{e}_{\bf k} + \int
\ik \frac{H' \delta_1 Q_{k}}{2\pi} \hat{e}_{\bf k_*-k}\,.
\end{equation} 
and hence we can identify the 
second order scalar field perturbation at Hubble exit:
\bea
\frac12 \delta_2 Q_{\bf k*}&=&
\frac{H'}{H}(\delta_1 Q \bullet \delta_1 Q)_{\bf k*}\,,
\label{dvphik}
\eea
where we have taken $\delta_1Q_{\bf k_*-k'}= (H/2\pi)\hat{e}_{\bf
  k_*-k'}$ for $k_*\gg k'$ and a ``$\bullet$'' represents a cut-off
convolution, that is a convolution-type integral defined by
\be
(f\bullet g)_{\bf k}\equiv\int^{k_c}_{0}d^3k'f_{\bf k'}g_{{\bf k}-{\bf k'}}\,,
\ee
where $k_c\ll k$.

%



%

This effect gives a non-zero bispectrum (\ref{defbispectrum}) for the
scalar field perturbations. In the squeezed triangle limit, $k_2\simeq
k_3$ and $k_1\ll k_2$, the correlation of the field perturbations at
the point when the two smaller modes cross the horizon ($k_2=aH$)
gives, to leading order,
\be
\langle\delta Q_{\bf k_1}\,\delta Q_{\bf k_2}\,\delta Q_{\bf k_3}\rangle
 = \langle \delta_1Q_{\bf k_1}\,\frac12 \delta_2Q_{\bf
   k_2}\,\delta_1Q_{\bf  k_3}\rangle
+ \langle \delta_1Q_{\bf k_1}\, \delta_1Q_{\bf
   k_2}\, \frac12 \delta_2Q_{\bf  k_3}\rangle
 \,,  
\label{dddvphi}
\ee
Note that the second order part of the perturbation on the larger
scale, $\delta_2Q_{\bf k_1}$, will be uncorrelated with the modes on
smaller scales.
Substituting our expression (\ref{dvphik}) for the second-order field
perturbation at horizon crossing into Eq.~(\ref{dddvphi}) yields
\bea
\langle\delta Q_{\bf k_1}\,\delta Q_{\bf k_2}\,\delta Q_{\bf
  k_3}\rangle
 &=&
 2\frac{H'}{H} \langle \delta_1Q_{\bf k_1}\, \delta_1Q_{\bf k_2}\,
 (\delta_1Q \bullet \delta_1Q)_{\bf k_3} \rangle\,,\\
&=& - 2\epsilon
 \left(\frac{H}{\dot\vphi}\right)
 P_Q( k_1) P_Q( k_2) \delbar{{\bf k_1+k_2+k_3}}
\eea
Comparing this with the definition of the bispectrum in
Eq.~(\ref{defbispectrum}) we see that
\be
 \label{BQ}
 B_Q(k,k_*) = - 2\epsilon \left(\frac{H}{\dot\vphi}\right)
 P_Q( k) P_Q( k_*) \,.  
\ee 
Note that this is not strictly of the local form given in
Eq.~(\ref{localB}) due to the asymmetry between long wavelengths,
$k_1$, and shorter wavelengths, $k_2$ and $k_3$. However $P(k)\gg
P(k_*)$ for $k\ll k_*$ as ${\cal P}_\zeta(k)\propto k^3P_\zeta(k)$ generated
during slow-roll inflation is nearly scale-invariant, and hence this
difference is not significant in the squeezed triangle limit.

One can verify that this coincides, in the limit $k_1\ll k_2,k_3$,
with the bispectrum for scalar field perturbations given by Seery and
Lidsey in Ref.~\cite{Seery2} for the case of a scale-invariant spectrum.


\subsection{Non-Gaussianity of curvature perturbations}

The power spectrum of the curvature perturbation due to inflaton
fluctuations during single field inflation is given by the first-order
relation~(\ref{defzeta1Q}). Substituting this into Eq.~(\ref{defP})
gives
\begin{equation}
P_\zeta(k) = \left( \frac{H}{\dot\vphi} \right)^2 P_Q(k) \,.
\end{equation}
The power spectrum of scalar field perturbations at Hubble exit is
given by Eq.~(\ref{Powerphi}) and thereafter the non-linear curvature
perturbation remains constant, on super-Hubble scales, for adiabatic
perturbations even after slow-roll inflation ends \cite{LW03}.
Thus for super-Hubble modes with $k\ll aH$ we can write the power
spectrum in terms of quantities at Hubble-exit
\begin{equation}
{\cal P}_\zeta (k) \equiv \frac{4\pi k^3}{(2\pi)^3} P_\zeta(k)
 = \left( \frac{H^2}{2\pi\dot\vphi} \right)_{k=aH}^2 \,.
\end{equation}
Slow-roll gives an almost scale-invariant power spectrum of
curvature perturbations on super-Hubble scales, whose tilt is given by
\begin{equation}
 \label{defn}
n - 1 \equiv \frac{d\ln{\cal P}_\zeta}{d\ln k} = -6\epsilon + 2\eta
\,.
\end{equation}

We can write the bispectrum (\ref{defbispectrum}) for the curvature
perturbation in terms of a local transformation between second-order
gauge-invariant variables, Eq.~(\ref{defzeta2Q2}), and the intrinsic
bispectrum of the scalar field perturbations. Substituting
Eqs.~(\ref{defzeta1Q}) and~(\ref{defzeta2Q2}) into Eq.~(\ref{defB}) we
obtain
\begin{eqnarray}
B_\zeta({\bf k_1},{\bf k_2}) = - \left( \frac{H}{\dot\vphi} \right)^3
 B_Q({\bf k_1},{\bf k_2}) + (2\epsilon-\eta) \left( P_\zeta(k_1)
   P_\zeta(k_2) + P_\zeta(k_2) P_\zeta(k_3) + P_\zeta(k_3)
   P_\zeta(k_1) \right) \,.
\end{eqnarray}
We see that if the field perturbations on spatially flat hypersurfaces
were strictly Gaussian at second order, $B_Q=0$, then the second
order curvature perturbation $\zeta_2$ would be of the ``local'' form
defined in Eq.~(\ref{deffnl}) with $f_{NL}=5(\eta-2\epsilon)/6$.

In the squeezed triangle limit, $k\ll k_*$, the scalar field
bispectrum at Hubble exit is given by (\ref{BQ}) and thus
\begin{equation}
 \label{Bzeta}
B_\zeta(k,k_*) = (6\epsilon-2\eta) P_\zeta(k) P_\zeta(k_*) \,,
\end{equation}
where we have used $P_\zeta(k)\gg P_\zeta(k_*)$ for a nearly
scale-invariant spectrum. Thus comparing with Eq.~(\ref{localB}) in this
squeezed triangle limit, we can identify \cite{Maldacena}
\begin{equation}
f_{NL} = \frac{5}{12} (-6\epsilon+2\eta) \,.
\end{equation}
As noted by several authors \cite{Creminelli,Gruzinov} this gives a
consistency relation between the primordial non-Gaussianity and the
tilt of the power spectrum (\ref{defn})
\begin{equation}
f_{NL} = \frac{5}{12} (n-1) \,.
\end{equation}
which does not rely on slow-roll during inflation, but does rely on the
adiabaticity of the perturbations on super-Hubble scales and thus the
existence of a conserved curvature perturbation after Hubble-exit.

\section{Multi-field inflation}

This discussion of of the non-Gaussianty of primordial curvature
perturbations from inflation can be readily extended to the
multi-field \cite{LRb} case by generalising Eq.~(\ref{defzeta})
\begin{equation}
\zeta = \sum_i N_{,i} \delta Q_i + \frac12 
\sum_{i,j} N_{,ij} \delta Q_i \delta Q_j + \ldots \,.
\end{equation}
where we identify $\zeta$ with the perturbed expansion of a uniform
density hypersurface with respect to an initial uniform-expansion
hypersurface during inflation, and we define $N_{,i}\equiv
dN/d\varphi_i$. 
Hence to first and second order we have
\begin{eqnarray}
\label{defzeta1Ni}
 \zeta_1 &=& \sum_i N_{,i} \delta_1 Q_i \,,\\
\label{defzeta2Nij}
 \zeta_2 &=& \sum_i N_{,i} \delta_2 Q_i + \sum_{i,j} N_{,ij} \delta_1 Q_i
 \delta_1 Q_j \,.
\end{eqnarray}
The gauge-invariant definition of field perturbations on uniform-expansion hypersurfaces, $\delta Q$, was given in Eqs.~(\ref{defQ1}) and (\ref{defQ2}). 

It is important to realise in the multi-field case
that the primordial curvature perturbation is evaluated on a uniform-density hypersurface at the 
some primordial epoch after inflation has ended. Thus the integrated expansion to this uniform-density hypersurface as a function of the initial field local values, $N(\vphi_i)$, must in general be calculated not only during inflation, but throughout the subsequent cosmological history, including
reheating. Only in the case of adiabatic perturbations on super-Hubble scales, where the perturbed evolution follows the background trajectory in phase space \cite{WMLL,Gordon}, can we equate $\delta N$ with the perturbed expansion at Hubble-exit, independently of the subsequent expansion history, as was done in Eqs.~(\ref{defzeta}--\ref{d2N}) for the single-field case.
Nonetheless, with this caveat we can formally write down the primordial perturbation in terms of the integrated expansion $N$ and it's dependence on the different fields.

Thus substituting Eqs.~(\ref{defzeta1Ni}) and (\ref{defzeta2Nij}) into the definitions of the power spectrum (\ref{defP}) and bispectrum (\ref{defB}), we obtain
\begin{eqnarray}
P_\zeta (k) &=& \sum_j N_{,i} N_{,j} \, C_{ij}(k) \,,\\
 \label{Bzetaijm}
B_\zeta(k_1,k_2,k_3) &=& \sum_{i,j,m} N_{,i} N_{,j} N_{,m}
B_Q^{i,j,m}(k_1,k_2,k_3) + \sum_{i,j,m,n} N_{,i} N_{,j} N_{,mn}
 \left[ C_{im}(k_1) C_{jn}(k_2) + {\rm perms} \right] \,,
\end{eqnarray}
where we define the multi-field cross-correlation and bispectrum at leading order
\begin{eqnarray}
 \label{defCij}
\langle \delta_1 Q_{i,{\bf k}} \delta_1 Q_{j,{\bf k'}} \rangle
 &\equiv& C_{ij}(k) \delbar{{\bf k}+{\bf k'}} \,,\\
 \label{defBQijm}
\langle\delta Q_{i,{\bf k_1}}\,\delta Q_{j,{\bf
    k_2}}\,\delta Q_{m,{\bf k_3}}\rangle
 &\equiv&
 B_Q^{i,j,m}(k_1,k_2,k_3)  \delbar{{\bf k_1}+{\bf k_2}+{\bf k_3}} \,.
\end{eqnarray}

Equation~(\ref{Bzetaijm}) allows for an arbitrary bispectrum, $B_Q^{i,j,m}$, for the field perturbations on spatially flat hypersurfaces and generalises the result of Lyth and Rodriguez \cite{LRb} to allow for correlations between the fields on super-Hubble scales.
In principle it does not rely on slow-roll, only that the subsequent expansion can be expressed as a function solely of the local field values. Nonetheless in the rest of this paper we will restrict our analysis to slow-roll in order to calculate the bispectrum of the field perturbations.

\subsection{Slow-roll field perturbations}

The field perturbations originate as vacuum fluctuations on
sub-Hubble scales and hence $\delta Q_i$ and $\delta Q_j$ are independent random
variables at Hubble-exit with
\begin{equation}
 \label{defQik}
\delta_1 Q_{i,{\bf k_*}} = \frac{H}{2\pi} \hat{e}_{i,{\bf k_*}} \,,
\end{equation}
to leading order in a slow-roll expansion, where 
\begin{equation}
\langle \hat{e}_{i,{\bf k}} \hat{e}_{j,{\bf k'}} \rangle
 = \frac{\delta_{ij}}{4\pi k^3} \delbar{{\bf k}+{\bf k'}} \,.
 \label{eiej}
 \end{equation}
At Hubble-exit we have $C_{ij}(k_*) = (H^2/2k_*^3) \delta_{ij}$, but
coupled evolution of linear perturbations on large scales
(proportional to $\partial^2 V/\partial\vphi_i\partial\vphi_j$ in
slow-roll approximation) can give rise to off-diagonal terms in
multi-field inflation \cite{Bartolo,Wands02}. Such off-diagonal terms are small in the slow-roll
approximation, but their effect can become significant when integrated over many expansion
times for long-wavelengths, $k\ll k_*$.

Once again we can estimate the bispectrum of the scalar field
perturbations in the squeezed triangle limit $k\ll k_*$ at Hubble-exit
$k_*=aH$, due solely to gravitational coupling in slow-roll
inflation. Generalising Eq.~(\ref{dvphik}) to multi-field inflation we
obtain
\bea
\frac12 \delta_2 Q_{i, \bf k*}
 &=& \sum\limits_j \frac{H_{,j}}{H} (\delta_1Q_j\bullet\delta_1Q_i)_{\bf k*}\,,
\eea
where $H_{,i}\equiv dH/d\vphi_i$.
Hence for $k_1\ll k_2\simeq k_3\simeq k_*$
%
\bea
\langle\delta Q_{i,{\bf k_1}}\,\delta Q_{j,{\bf k_2}}\,\delta Q_{m,{\bf
  k_3}}\rangle
 &=& \sum_n
 \frac{H_{,n}}{H} \left( \langle \delta_1Q_{i,{\bf k_1}}\, (\delta_1Q_n \bullet
 \delta_1Q_j)_{\bf k_2} \, \delta_1Q_{m,{\bf k_3}}
 \rangle
 + \langle \delta_1Q_{i,{\bf k_1}}\, \delta_1Q_{j,{\bf k_2}}\,
 (\delta_1Q_n \bullet \delta_1Q_m)_{\bf k_3} \rangle \right)
\,,\nonumber \\
&=& 
\label{QiQjQm}
 \sum_n 2 \frac{H_{,n}}{H} C_{in}(k_1) C_{jm}(k_2) \delbar{{\bf
    k_1+k_2+k_3}} \,.
\eea
%

Comparing (\ref{QiQjQm}) with the definition of the multi-field bispectrum in
Eq.~(\ref{defBQijm}) we see that in the squeezed triangle limit $k\ll
k_*$ at Hubble-exit
\be
 \label{BQijm}
B_Q^{i,j,m}(k,k_*) = \sum_n 2 \frac{H_{,n}}{H} C_{in}(k) C_{jm}(k_*) \,.  
\ee
As we would expect, the bispectrum for field fluctuations due to gravitational coupling is non-zero only when the Hubble rate at Hubble-exit is dependent upon the field values.

Following Ref.~\cite{Gordon} we can use the unperturbed background solution, $\vphi_i(t)$, to define the instantaneous adiabatic or ``inflaton'' perturbation lying along the background trajectory in multi-field inflation.
\begin{equation}
 d\sigma \equiv \frac{\dot\vphi_i}{\sqrt{\sum_j\dot\vphi_j^2}} d\vphi_i \,.
\end{equation}
All orthogonal directions in field-space describe instantaneous isocurvature or entropy field perturbations~\cite{Gordon,RMP} which we denote generically by $\delta\chi$.

In the slow-roll approximation the background trajectory is given by the gradient of the potential, or equivalently the Hubble rate. Thus
 \begin{equation}
 \sum_n \frac{H_{,n}}{H} \delta_1 Q_n = \frac{H_{,\sigma}}{H} \delta_1 Q_\sigma
   \,.
 \end{equation}
Thus the bispectrum (\ref{BQijm}) is only non-zero for long-wavelength field perturbations $\delta Q_i$ which are correlated with the instantaneous inflaton direction, $\sigma$:
\be
B_Q^{i,j,m}(k,k_*) = -2\epsilon \frac{H}{\dot\sigma} C_{i\sigma}(k) C_{jm}(k_*)
 \,,
\ee
analogous to the single-field case (\ref{BQ}), where the slow-roll parameter (\ref{defepsilon}) is given by
\begin{equation}
 \epsilon = - \frac{dH}{d\sigma} \frac{\dot\sigma}{H^2} \,.
\end{equation}

Isocurvature field perturbations that remain decoupled from, and hence uncorrelated with, the inflaton perturbation (i.e., $C_{\chi\sigma}(k)=0$), retain a purely Gaussian distribution with $B_\chi(k,k_*)=0$ unless we consider non-gravitational interactions, neglected here in our slow-roll approximation.

\subsection{Non-Gaussianity of curvature perturbation in slow-roll}

Finally we can write the bispectrum (\ref{Bzetaijm}) for the curvature perturbation for multi-field slow-roll inflation in the squeezed triangle limit, $k\ll k_*$, as
\begin{eqnarray}
 \label{Bzetasigmajm}
B_\zeta(k,k_*) &=& 2 \sum_{i,j} N_{,i} N_{,j} \left( \epsilon N_{,j} N_{,\sigma} C_{i\sigma}(k) + \sum_n N_{,jn} C_{in}(k) \right) P_{Q_j}(k_*) \,.
\end{eqnarray}
One verify that this reduces to Eq.~(\ref{Bzeta}) in the single field case where we have $N_{,\sigma}$ and $N_{,\sigma\sigma}$ given by Eqs.~(\ref{d1N}) and (\ref{d2N}). This single-field result for the non-Gaussianity still holds in a multi-field setting so long as the expansion history is independent of all the other fields.

An alternative limit is the case where the expansion history is dominated by an isocurvature field, such that $|N_{,\chi}|\gg |N_{,\sigma}|$. This would apply for instance to the curvaton scenario \cite{curvaton}. In this case the primordial curvature perturbation is given by
\begin{equation}
 P_\zeta (k) = (N_{,\chi})^2 P_\chi (k) \,.
\end{equation}
There is no non-Gaussianity due to gravitational coupling, $B_\chi=0$, for an isocurvature field that remains decoupled from the  from the inflaton, so that $C_{\chi\sigma}=0$. Hence the non-Gaussianity of the primordial curvature perturbation (\ref{Bzetaijm}) is given by
\begin{eqnarray}
 B_\zeta(k_1,k_2,k_3) &=& (N_{,\chi})^2 N_{,\chi\chi}
 \left[ P_\chi(k_1) P_\chi(k_2) + {\rm perms} \right]
  \,,
  \nonumber\\
  &=& \frac{N_{,\chi\chi}}{(N_{,\chi})^2} \left[ P_\zeta(k_1) P_\zeta(k_2) + {\rm perms} \right] \,.
\end{eqnarray}
This is purely of the local form (\ref{localB}) and we can identify \cite{LRb}
\begin{equation}
f_{NL} = -\frac{5}{6} \frac{N_{,\chi\chi}}{(N_{,\chi})^2} \,.
\end{equation}

Even in the general case it can be shown that the contribution of the intrinsic non-Gaussianity of the field perturbations to the bispectrum of the curvature perturbations must be small in the slow-roll approximation \cite{LZ}.

\section{Conclusions}

We have shown how the non-Gaussianity of primordial curvature perturbations is described using gauge-invariant cosmological perturbations at second-order. The split between zeroth, first and second-order parts of inhomogeneous fields is arbitrary, however if we expand about a FRW cosmology, the zeroth-order part of cosmological fields can be taken to be spatially homogeneous. In this paper we have further assumed that the first-order part can be taken to be a Gaussian random field, and the second-order part describes the non-linear evolution of inhomogeneities, and hence the non-Gaussian part of the distribution.

The arbitrariness of the choice of coordinates leaves a residual gauge-freedom. Nonetheless one can construct gauge-invariant quantities, at any order, by specifying an unambiguous, physical quantity \cite{MW}. In particular we have given a manifestly gauge-invariant definition of $\zeta$, the perturbed expansion on uniform density hypersurfaces on large scales, used by many authors to describe the primordial metric perturbation as it remains constant, at any order, for adiabatic perturbations \cite{LW03,Rigopoulos1,LMS,LangloisVernizzi}. This is related to the dimensionless density perturbation on spatially flat hypersurfaces, however the non-linear relation between the two leads to a non-Gaussian metric perturbation for a purely Gaussian density perturbation, and vice versa.

In the case of single-field inflation, the field perturbations become adiabatic in the large-scale limit \cite{Vernizzi} and the primordial $\zeta$ long after inflation has ended can be equated with the perturbed expansion on uniform-field hypersurfaces at Hubble exit. We have shown how this is equivalent to the non-linear extension of the $\delta N$-formalism \cite{Starobinsky,SasakiStewart,LMS,LRb} which allows one to calculate the primordial perturbation in terms of $\delta Q$, the field perturbations on spatially flat hypersurfaces, at Hubble exit and the unperturbed FRW solution for the integrated expansion, $N(\varphi)$. This formalism is especially compact for multiple fields where the primordial metric perturbation (say, during primordial nucleosynthesis) may be completely different from that at Hubble exit during inflation. On the other hand one should bear in mind that $N(\varphi_i)$ is a non-linear function of several scalar fields which must be determined by solving the background field equations to determine the integrated expansion up to some fixed primordial density as a function of the field values during inflation. Nonetheless Lyth and Rodriguez \cite{LRb} have recently shown how powerful this method is for describing the non-Gaussianity of the primordial perturbation due to non-linear evolution on super-Hubble scales. 

In single- or multi-field inflation one can give the leading-order expression for the bispectrum of the primordial metric perturbation in terms of the power spectrum and the intrinsic bispectrum of the fields at Hubble-exit during inflation. 
We have used a simple argument to estimate the intrinsic bispectrum of the field perturbations at Hubble exit due to gravity in the limit of degenerate triangles during slow-roll inflation with single or multiple fields, and hence give an expression for the non-linearity of the primordial density perturbations from slow-roll inflation in this limit. In line with previous results we find that the non-Gaussianity of field perturbations is suppressed during slow-roll, and we have shown that it vanishes at leading order for isocurvature perturbations during inflation. 
Our final result in Eq.~(\ref{Bzetasigmajm}) generalises earlier work to allow for non-scale invariance of the scalar field power spectra and correlations between the fields that can develop on super-Hubble scales.

\section*{Acknowledgements}

The authors are grateful to Karim Malik for useful comments.
This work was supported by a PPARC studentship and PPARC grant PPA/G/S/2000/00115.

\end{document}